\documentclass[aps,reprint,preprintnumbers,amsmath,amssymb,superscriptaddress]{revtex4-1}

\usepackage{graphicx,color}

\usepackage{dcolumn}

\usepackage{bm}

\usepackage[dvipdfmx]{hyperref}

\begin{document}

\title{Three-dimensional Bose-Einstein Condensation\\in the Spin-1/2 Ferromagnetic-leg Ladder 3-Br-4-F-V}

\author{Y. Kono}
\email{k-yohei@issp.u-tokyo.ac.jp}
\affiliation{Institute for Solid State Physics, the University of Tokyo, Kashiwa 277-8581, Japan}
\author{H. Yamaguchi}
\affiliation{Department of Physical Science, Osaka Prefecture University, Osaka 599-8531, Japan}
\author{Y. Hosokoshi}
\affiliation{Department of Physical Science, Osaka Prefecture University, Osaka 599-8531, Japan}
\author{T. Sakakibara}
\affiliation{Institute for Solid State Physics, the University of Tokyo, Kashiwa 277-8581, Japan}

\date{\today}

\begin{abstract}
The critical exponent of the phase boundary has been examined on the three-dimensional incommensurate ordering phase in the spin-1/2 ferromagnetic-leg ladder 3-Br-4-F-V [=3-(3-bromo-4-fluorophenyl)-1,5-diphenylverdazyl]. Using the temperature-window fitting technique, we obtained the critical exponents which agreed with the three-dimensional (3D) Bose-Einstein condensation (BEC) universality class at both sides of the lower critical field and the saturation field. 3-Br-4-F-V thus becomes a new member of the quantum magnets which prove the universality of the 3D BEC exponent.
\end{abstract}



\maketitle

\section{Introduction}
Quantum phase transitions (QPTs), arising from quantum fluctuations, are characterized by critical exponents near quantum critical points (QCPs): observables such as specific heat, magnetic susceptibility, and correlation function show characteristic power laws in the temperature dependencies with universal critical exponents near QCPs~\cite{2011qpbookS,0034-4885-66-12-R01}. Since the universality stems only from dimensionality of the system and symmetries of order parameters, the determination of the critical exponents is of fundamental importance for understanding the nature of QPTs.

For the classification of magnetic-field-induced QPTs, so-called “Bose-Einstein condensation (BEC) of magnons” has been an attractive topic in condensed-matter physics for the past few decades; since the exact mapping between a lattice gas of hard-core bosons and a spin-1/2 system was first introduced in 1956~\cite{Matsubara01121956}, the notion of a BEC has been extended to the quantum spin systems and its possibility has been investigated in a large number of model compounds~\cite{giamarchi2008bose,RevModPhys.86.563}. 
For the realization of a BEC in quantum magnets, uniaxial [U(1)] symmetry of the original spin Hamiltonian is required in order to satisfy a number conservation of bosons. 
A spontaneous breaking of the symmetry at a QCP then becomes a BEC QCP. 
A promising realization of this situation is a weakly coupled spin-dimer system that exhibits a three-dimensional (3D) $XY$ antiferromagnetic (AFM) ordering in a magnetic field~\cite{RevModPhys.86.563}. 
In such systems, field-induced QCPs exist at two critical fields, $H_{\mathrm{c1}}$ and $H_{\mathrm{c2}}$. A spin gap of a quantum disordered state is destroyed at $H_{\mathrm{c1}}$, and a full saturation of the magnetization occurs at the upper critical field $H_{\mathrm{c2}}$. 
Both of these quantum phase transitions belong to the 3D BEC universality class, and the critical exponent $\nu$ of the phase boundary, defined by $T\sim\left|H_{\mathrm{c1,2}}(T)-H_{\mathrm{c1,2}}(0)\right|^{\nu}$, has been predicted to be  $\nu\,=\,2/3$ as $T\,\rightarrow\,0$~\cite{PhysRevB.59.11398,PhysRevLett.84.5868}. 

Typical experimental examples in which the 3D BEC exponent at $H_{\mathrm{c1}}$ has been confirmed include spin dimer systems TlCuCl$_{3}$~\cite{JPSJ.77.013701} and BaCuSi$_{2}$O$_{6}$~\cite{PhysRevB.72.100404,*sebastian2006dimensional}.
Unfortunately in these systems, however, the test of BEC at $H_{c2}$ has been lacking because the saturation fields in such systems are usually in excess of tens of teslas.

\begin{figure}[b]
\begin{center}
\includegraphics[width=0.7\linewidth]{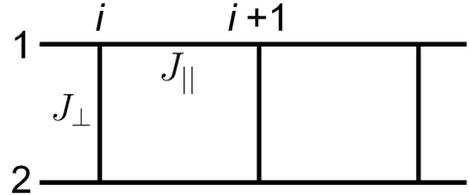}
\caption{Schematic picture of a two-leg ladder described by Eq.~(\ref{ladder}).}\label{f1}
\end{center}
\end{figure}

Another interesting spin gapped system, yet less studied in the context of BEC, is a spin-1/2 two-leg ladder. The Hamiltonian of the system, in the simplest form, can be expressed as
\begin{eqnarray}
\mathcal{H}&=&J_{||}\sum_{i,\alpha}\bm{S_{i,\alpha}\cdot S_{i+1,\alpha}}+J_{\perp}\sum_{i}\bm{S_{i,1}\cdot S_{i,2}}\nonumber\\&&-g\mu_{B}H\sum_{i,\alpha}S^{z}_{i,\alpha},\label{ladder}
\end{eqnarray}
where $J_{||}$ is the interaction along each leg ($\alpha=1,2$), $J_{\perp}$ is the rung interaction between the legs, $g$ is the $g$ factor, and $\mu_{B}$ is the Bohr magneton (Fig.~\ref{f1}). 

The most frequently studied case is ``AFM-AFM''  ($J_{||}$, $J_{\perp}\,>\,0$), to which the compounds (Cu$_{7}$H$_{10}$N)$_{2}$CuBr$_{2}$ (DIMPY)~\cite{PhysRevLett.108.097201} and (Cu$_{5}$H$_{12}$N)$_{2}$CuBr$_{4}$ (BPCB) correspond~\cite{PhysRevLett.101.247202}. 
In this case, the ground state is a singlet state ($S\,=\,0$), and there always exists a spin gap~\cite{PhysRevLett.77.1865}. 
With increasing the magnetic field, the lowest branch of triplet states ($S_{z}=1$) degenerates into the singlet state at $H_{c1}$, and $S_{z}=1$ bosons (triplons) are excited. 
Consequently, a Tomonaga-Luttinger liquid (TLL) state appears in the gapless phase between $H_{c1}$ and $H_{c2}$ due to the one dimensionality~\cite{PhysRevB.59.11398,Giamarchi}. If there exist 3D interactions between the ladders---as is usual in real magnets---the triplons can condense into the 3D BEC state~\cite{PhysRevB.59.11398}. 
There have been, however, few experimental tests for the 3D BEC exponent in this case because of a large spin gap over tens of teslas, e.g., SrCu$_{2}$O$_{3}$~\cite{PhysRevLett.73.3463}, or a low 3D ordering temperature less than 100\,mK, e.g., BPCB~\cite{PhysRevLett.101.247202}.

Recently, spin-1/2 ferromagnetic-leg (FM-leg) ladders ($J_{||}\,<\,0$, $J_{\perp}\,>\,0$) have been synthesized for the first time, using verdazyl radical molecules~\cite{PhysRevLett.110.157205,doi:10.7566/JPSJ.83.033707,PhysRevB.89.220402,PhysRevB.91.125104}. Theoretically, FM-leg ladders with an isotropic leg interaction also have  a spin gap, which stems from the rung-singlet state~\cite{PhysRevB.53.R8848,*PhysRevB.67.064419,*PhysRevB.72.014449}, and
the ground state between $H_{c1}$ and $H_{c2}$ has also been predicted to be a TLL~\cite{PhysRevB.70.014425}. A 3D BEC state is thus expected to be induced in the intermediate field range by weak 3D interactions. 
An advantage of the FM-leg ladders over the AFM-AFM ones is that they provide more opportunities to access the upper QCP at $H_{c2}$; because $H_{c2}$ is insensitive to FM interactions~\cite{PhysRevB.70.014425,PhysRevB.75.134421}, the FM-leg case gives smaller $H_{c2}$ when the intraladder couplings are of the same order of magnitude.

Among the three FM-leg ladders synthesized to date, 3-Br-4-F-V [=3-(3-bromo-4-fluorophenyl)-1,5-diphenylverdazyl] has the largest rung interaction ($J_{||}=-8.3$~K, $J_\perp=12.5$~K, $|J_\perp/J_{||}|=1.5$) and only this compound has a spin gap ($\sim$\,5\,T)~\cite{doi:10.7566/JPSJ.83.033707,PhysRevB.89.220402}. The other two compounds are antiferromagnetically ordered in the ground state due to 3D interactions~\cite{doi:10.7566/JPSJ.83.033707}. 
Compared with BPCB~\cite{PhysRevLett.101.247202}, which has a similar rung-coupling constant ($J_{||}/k_{B}$\,=\,3.3\,K, $J_{\perp}/k_{\mathrm{B}}$\,=\,12.9\,K, and $\mu_{0}H_{\mathrm{c2}}\sim$\,14.5\,T), 3-Br-4-F-V actually has the smaller saturation field, $\mu_{0}H_{\mathrm{c2}}\sim$\,9\,T. In the magnetic field range between $H_{\mathrm{c1}}$ and $H_{\mathrm{c2}}$, 3-Br-4-F-V exhibits a 3D incommensurate ordering due to interladder couplings below $\sim$0.6~K as revealed by nuclear magnetic resonance (NMR) measurements~\cite{PhysRevB.89.220402}. A TLL region was expected to exist above the 3D ordering temperatures as is the case for (quasi-)1D gapped spin systems~\cite{maeda2007universal,PhysRevLett.108.097201,PhysRevLett.101.247202}, but, in reality, most of the anticipated TLL region is replaced by a partial ordering phase in 3-Br-4-F-V~\cite{PhysRevB.89.220402}. 
Although detailed magnetic structures of these exotic ordering phases remain to be clarified, a 3D BEC is expected to be realized in this situation.
The 3D ordering phase is indeed domelike, in close resemblance to the other model compounds such as BaCuSi$_{2}$O$_{6}$~\cite{PhysRevB.72.100404,*sebastian2006dimensional}.
Moreover, spin couplings between verdazyl radical molecules are essentially isotropic~\cite{doi:10.7566/JPSJ.83.033707}, in favor of U(1) symmetry of its spin Hamiltonian.

In this paper, we focus on the critical exponents of the 3D ordering phase boundary in 3-Br-4-F-V. 
The phase boundary near $H_{\mathrm{c1}}$ and $H_{\mathrm{c2}}$ is precisely determined by means of specific-heat and dc magnetization measurements. Applying the temperature-window technique~\cite{PhysRevLett.95.127202,*PhysRevLett.96.189704,PhysRevB.72.100404,JPSJ.77.013701}, the critical exponents consistent with the 3D BEC scenario are obtained at \emph{both sides} of $H_{\mathrm{c1}}$ and $H_{\mathrm{c2}}$ at low temperatures.

\section{Experimental}
Single-crystal samples of 3-Br-4-F-V were synthesized as described in Ref.~\onlinecite{doi:10.7566/JPSJ.83.033707}.
Specific-heat measurements were carried out by the standard quasiadiabatic heat-pulse and relaxation methods on a 1.46-mg sample (\#1). Direct-current magnetization measurements were performed by a force magnetometer~\cite{JJAP.33.5067} on a 7.42-mg sample (\#2), taken from the same batch. In both measurements, a $^{3}$He-$^{4}$He dilution refrigerator was used in the temperature ranges 0.1\,K\,$\le$\,$T$\,$\le$\,1\,K. 
In all the measurements, magnetic fields up to 9\,T were applied perpendicular to the $\bm{a}$ axis (perpendicular to the leg direction).  

\section{Results and Discussion}
\subsection{Temperature dependence of the specific heat}
\begin{figure}[b]
\begin{center}
\includegraphics[width=0.95\linewidth]{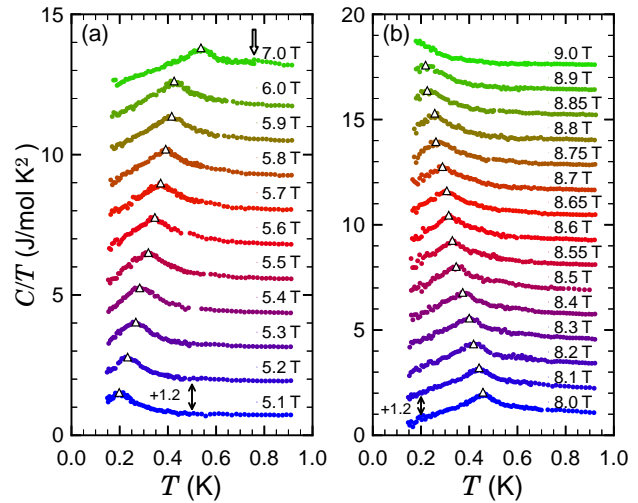}
\caption{Temperature dependence of the magnetic heat capacity in several magnetic fields near (a) the lower critical field $H_{\mathrm{c1}}$ and (b) the upper critical (saturation) field $H_{\mathrm{c2}}$. Each curve is shifted by $+1.2$\,J/mol K$^2$ for clarity. Open triangles denote the peak positions indicating the 3D ordering temperatures, $T_{c}$. The open arrow denotes the partial-ordering temperature at 7\,T.}\label{f2}
\end{center}
\end{figure}

In Figs.~\ref{f2}(a) and \ref{f2}(b), we show the heat capacity divided by temperature, $C/T$, in several magnetic fields near $H_{c1}$ and $H_{c2}$, respectively. 
In these data, nuclear Schottky contributions from $^{1}$H, $^{19}$F, and $^{14}$N are subtracted. 
A sharp peak indicative of the 3D ordering can be observed. On the other hand, the shoulderlike anomaly inferred in the previous report~\cite{PhysRevB.89.220402} to be a partial ordering is weak (only the one at 7\,T is indicated by an open arrow) or even indiscernible within the experimental resolution. This weakness of the partial-ordering anomaly might be due to a difference of the sample quality from the previous one.

\begin{figure}[t]
\begin{center}
\includegraphics[width=0.95\linewidth]{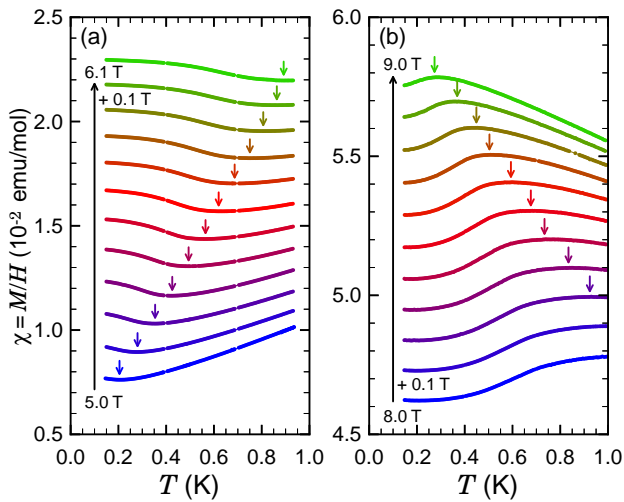}
\caption{Temperature dependence of the magnetic susceptibility $\chi\,=\,M/H$ in several magnetic fields (a) from 5\,T to 6.1\,T and (b) from 8\,T to 9\,T in 0.1\,T steps. Arrows indicate the temperature $T_{\mathrm{ex}}$ at which $\chi$ takes the nontrivial (a) minimum or (b) maximum.}\label{f3}
\end{center}
\end{figure}

\begin{figure}[t]
\begin{center}
\includegraphics[width=0.95\linewidth]{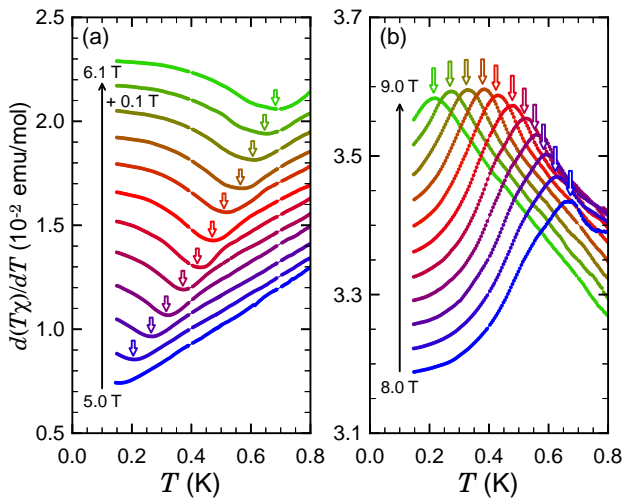}
\caption{Temperature derivative of $T\chi$ [$d(T\chi)/dT$] obtained from Figs.~\ref{f3}(a) and \ref{f3}(b). Open arrows indicate the temperature at which $d(T\chi)/dT$ takes the nontrivial (a) minimum or (b) maximum.}\label{f4}
\end{center}
\end{figure}

\subsection{Temperature dependence of the magnetic susceptibility}
Figure \ref{f3} shows the temperature dependence of the magnetic susceptibility $\chi\,=\,M/H$ in several magnetic fields near the critical fields $H_{c1}$ and $H_{c2}$. There exists a nontrivial minimum (maximum) in Fig.~\ref{f3}(a) [Fig.~\ref{f3}(b)] (solid arrows), and the temperature at which the extremum appears ($T\,=\,T_{\mathrm{ex}}$) increases (decreases) with increasing the magnetic field. Analogous behavior in $\chi$ has been observed in typical spin-1/2 two-leg spin ladders such as DIMPY~\cite{PhysRevLett.108.097201}. Theoretically, such extremum in (quasi-)1D gapped spin systems can be interpreted as a crossover to the low-temperature TLL region~\cite{maeda2007universal}. 
Then, a 3D ordering exists slightly below $T_{\mathrm{ex}}$~\cite{PhysRevB.89.220402}.
In a previous report~\cite{PhysRevB.89.220402}, 3D ordering temperature, $T_{c}$, from $\chi$-$T$ was determined from a kink anomaly of the temperature derivative of $\chi$, and it was in good agreement with $T_{c}$ determined from a peak of the temperature dependence of the specific heat $C$. This correspondence is considered to be associated with Fisher\rq{}s relation for antiferromagnets,
\begin{equation}
C(T)\sim a\frac{\partial}{\partial T}\left(T\chi\right), \label{eq:fisher}
\end{equation}
which describes that the specific heat near a second-order AFM phase transition can be scaled with the temperature derivative of $T\chi$ (the coefficient $a$ is a slowly varying function near the transition)~\cite{Fisher}. This relation has been confirmed experimentally in several materials, e.g., Ref~\onlinecite{PhysRevB.7.4197}. In this sense, it would be the most plausible way to define $T_{c}$ of this material from $d(T\chi)/dT$.

As can be seen from Fig.~\ref{f4}, the temperature derivative of $T\chi$ [$d(T\chi)/dT$] exhibits a dip [Fig.~\ref{f4}(a)] or peak [Fig.~\ref{f4}(b)] anomaly below $T_{\mathrm{ex}}$. Here, we define the temperature at which these anomalies exist as the 3D ordering temperatures ($T\,=\,T_{c}$). Note that the most singular part in Eq.~(\ref{eq:fisher}) arises from the temperature derivative of $\chi$ itself. Therefore, this definition of $T_{c}$ is essentially the same as the one used in the previous experiment~\cite{PhysRevB.89.220402}.

\subsection{Phase boundary}
\begin{figure}[t]
\begin{center}
\includegraphics[width=0.95\linewidth]{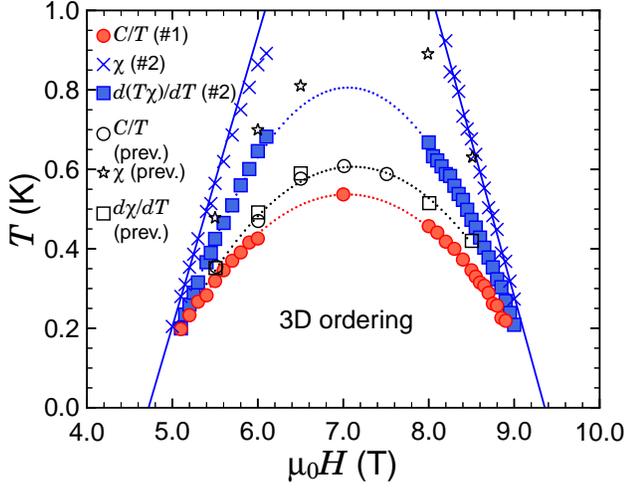}
\caption{Phase boundaries of the 3D ordering determined from the present measurements. Closed circles and squares show $T_{c}$ from the peak positions of $C/T$ (Fig.~\ref{f2}) and the dip or peak anomalies of $d(T\chi)/dT$ (Fig.~\ref{f4}), respectively. Open circles and squares denote previous (Ref.~\onlinecite{PhysRevB.89.220402}) results of $T_{c}$ from $C/T$ and $d\chi/dT$, respectively. Crosses (stars) denote the temperature $T_{\mathrm{ex}}$ at which the magnetic susceptibility $\chi$ takes extremum in the present (previous) results, and solid lines are a linear fit of the present ones at below 5.3\,T and above 8.6\,T. All the dotted lines are guides for the eye.}\label{f5}
\end{center}
\end{figure}

The 3D ordering temperatures observed in the present measurements are summarized in Fig.~\ref{f5} together with the results reported previously~\cite{PhysRevB.89.220402}. A noticeable difference between the present and the previous results is a disparity in the phase boundaries determined from the magnetization and specific-heat measurements;
whereas in the previous study these two measurements yielded nearly the same transition temperature, the phase boundary defined from $d(T\chi)/dT$ in the present experiment is rather higher in temperature than the one derived from $C/T$.
We ascribe this disparity to a high sensitivity of the interladder couplings to a strain in this system;
the larger crystal (sample \#2) used in the magnetization measurements shows slightly stronger interlayer couplings. The difference from the previous results might also be due to such sample dependence. 
We also note that the anomaly in $d(T\chi)/dT$ in Fig.~\ref{f4} (b) is broader than that in $d\chi/dT$ reported in the previous experiment~\cite{PhysRevB.89.220402}.
Considering these, we assess that the phase boundary determined from $C/T$ is more reliable in this study. Thus, in the next section, analysis of the critical exponent mainly focuses on the results from $C/T$, and uses the $d(T\chi)/dT$ data to support the results.

Another remarkable feature in Fig.~\ref{f5} is the linear behavior of $T_{\mathrm{ex}}$ against the magnetic field near $H_{\mathrm{c1}}$ and $H_{\mathrm{c2}}$. 
This behavior is reminiscent of a crossover to the TLL region in (quasi-)1D gapped spin systems~\cite{maeda2007universal}. 
Close to the critical field $H = H_{c}$, the crossover temperature $T_{\mathrm{ex}}$, at which the magnetization takes an extremum, is asymptotically expressed in the universal form
\begin{equation}
T_{\mathrm{ex}}(H)=c\,\frac{g\mu_{\mathrm{B}}}{k_{\mathrm{B}}}\left|H-H_{\mathrm{c1,2}}\right|,\label{tm}
\end{equation}
where the coefficient $c\,=\,0.762\,38$, and $k_{\mathrm{B}}$ is the Boltzmann constant. 
This linear relation has actually been observed on DIMPY~\cite{PhysRevLett.108.097201} for the lower critical field, and the ideal spin-1/2 one-dimensional AFM chain CuPzN~\cite{PhysRevLett.114.037202} for the saturation field. 
In the present case of  3-Br-4-F-V, a linear fit of the $T_{\mathrm{ex}}$ vs. $H$ plot yields the absolute value of the linear coefficient 0.73$\pm$0.01 at fields below 5.3\,T and 0.78$\pm$0.02 at fields above 8.6\,T, both of which are smaller than the value determined from Eq.~(\ref{tm}), $cg\mu_{\mathrm{B}}/k_{\mathrm{B}}\,=\,1.0246$, assuming $g=2$. 
Considering the fact that $T_{\mathrm{ex}}$ is relatively close to $T_{c}$ compared with the above example compounds, it seems that the temperature region at which Eq.~(\ref{tm}) can be applied is largely overlapped with the 3D ordering phase in 3-Br-4-F-V, although the linear relation between $T_{\mathrm{ex}}$ and $H$ can be ascribed to the quasi-one-dimensionality of this compound. In fact, such behavior has been observed in other quasi-1D quantum magnets~\cite{PhysRevB.91.060407,PhysRevB.95.020408}.

\begin{figure}[t]
\begin{center}
\includegraphics[width=\linewidth]{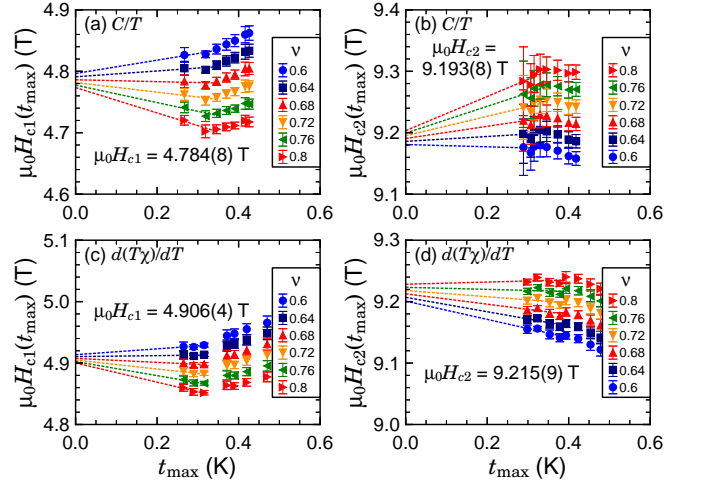}
\caption{Fitting results of the temperature-window technique with Eq.~(\ref{fit}), applied to the 3D ordering phase boundary. As a function of the maximum temperature of the window, $t_{\mathrm{max}}$, for several fixed $\nu$\rq{}s, (a) $H_{\mathrm{c1}}$ and (b) $H_{\mathrm{c2}}$ from $C/T$, and (c) $H_{\mathrm{c1}}$ and (d) $H_{\mathrm{c2}}$ from $d(T\chi)/dT$ are shown. The dotted lines are linear fitting lines to a part of the data on each $\nu$, used to determine $H_{\mathrm{c1,2}}(0)$ (see text).}\label{f6}
\end{center}
\end{figure}

\subsection{Fitting method to extract the critical fields and exponents of the phase boundary, and its results}
To extract the critical exponent $\nu$ of the 3D ordering phase boundary, we employ the temperature-window technique~\cite{PhysRevLett.95.127202,*PhysRevLett.96.189704,PhysRevB.72.100404,JPSJ.77.013701} in fitting the data by the power-law function $T\sim\left|H_{\mathrm{c1,2}}(T)-H_{\mathrm{c1,2}}(0)\right|^{\nu}$. Fortunately, the almost symmetric dome like phase diagram---in the boson language, this implies that the effect of quantum fluctuations on the boson effective mass at fields below $H_{\rm c1}$ is weak in this material~\cite{RevModPhys.86.563}---enables us to replace the function with a quadratic form   
\begin{equation}
T(h)=a\left(1-h^{2}\right)^{\nu},\label{fit}
\end{equation}
by which the critical field can be determined more accurately~\cite{PhysRevB.72.100404}. In Eq.~(\ref{fit}), $a$ is a fitting coefficient, which is approximated as a constant, and the normalized field is defined as $h\,=\,[H_{\mathrm{c1,2}}(T)-H_{m}]/|H_{m}-H_{\mathrm{c1,2}}(0)|$, where $H_{m}$ is the magnetic field centered in the 3D ordering dome (in this case, $\mu_{0}H_{m}\,\sim\,7$\,T).

We first fit Eq.~(\ref{fit}) to the data within the temperature window $0\,\le\,T\,\le\,t_{\mathrm{max}}$ for several fixed $\nu$\rq{}s, where $t_{\mathrm{max}}$ is the maximum temperature of the window. 
The lowest value of $t_{\mathrm{max}}$ used is 0.27\,K and 0.28\,K near $H_{\mathrm{c1}} (h=-1)$ and $H_{\mathrm{c2}} (h=1)$, respectively, so that at least three data points are available for the fitting. For the given $t_{\mathrm{max}}$, the fitting parameters $H_{\mathrm{c1,2}}(0)$ are determined [referred to as $H_{\mathrm{c1,2}}(t_{\mathrm{max}})$].
Iterating this procedure with increasing $t_{\mathrm{max}}$, we obtain $H_{\mathrm{c1,2}}(t_{\mathrm{max}})$ as a function of $t_{\mathrm{max}}$~\cite{note1}.
The results of $H_{\mathrm{c1}}(t_{\mathrm{max}})$ and $H_{\mathrm{c2}}(t_{\mathrm{max}})$ determined from $C/T$ are shown in Figs.~\ref{f6}(a) and \ref{f6}(b).
The zero-temperature limit $H_{\mathrm{c1,2}}(0)$ for each $\nu$ can be obtained by a linear extrapolation of the lowest few data points of $H_{\mathrm{c1,2}}(t_{\mathrm{max}})$. 
As shown in Figs.~\ref{f6}(a) and \ref{f6}(b), the linear fitting lines on each panel (dashed lines) converge on the narrow field region for a range of the $\nu$ values, analogous to the results of other compounds to which the same technique has been applied~\cite{PhysRevB.72.100404,PhysRevLett.95.127202,PhysRevLett.96.077204}. 
Thus, the critical fields $H_{\mathrm{c1,2}}(0)$ can be estimated to be $\mu_{0}H_{\mathrm{c1}}\,=\,4.784(8)$\,T for Fig.~\ref{f6}(a), and $H_{\mathrm{c2}}(t_{\mathrm{max}})\,=\,9.193(8)$\,T for Fig.~\ref{f6}(b), irrespective for the particular $\nu$ value. The same analysis is also applied to the $d(T\chi)/dT$ results and yields $\mu_{0}H_{\mathrm{c1}}\,=\,4.906(4)$\,T for Fig.~\ref{f6}(c) and $\mu_{0}H_{\mathrm{c2}}\,=\,9.215(9)$\,T for Fig.~\ref{f6}(d). Slight differences of the critical fields are attributed to a difference in the sample quality used for the magnetization measurements.

\begin{figure}[t]
\begin{center}
\includegraphics[width=0.97\linewidth]{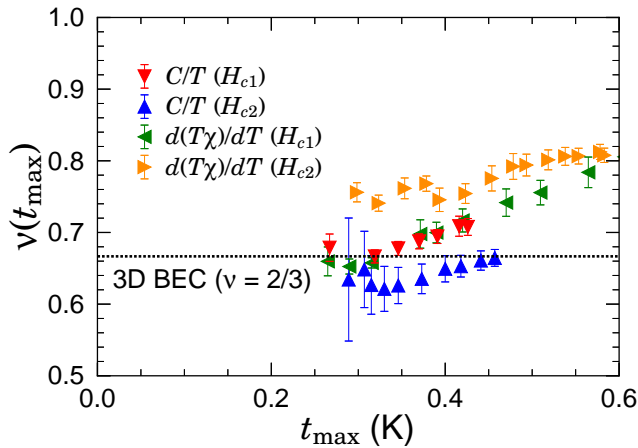}
\caption{Critical exponent $\nu$ as a function of $t_{\mathrm{max}}$ with fixed $H_{\mathrm{c1,2}}(0)$ determined from Fig.~\ref{f6}. Down and up triangles denote the results from $C/T$ near $H_{\mathrm{c1}}$ and $H_{\mathrm{c2}}$, respectively. Left and right triangles denote the results from $d(T\chi)/dT$ near $H_{\mathrm{c1}}$ and $H_{\mathrm{c2}}$, respectively. The dotted line shows the critical exponent at the QCP that belongs to the 3D BEC universality class ($\nu\,=\,2/3$). All the error bars represent fitting errors.}\label{f7}
\end{center}
\end{figure}

Fixing $H_{\mathrm{c1,2}}(0)$ to the values given above, the fitting of Eq.~(\ref{fit}) to the data, in a similar manner but now $\nu$ being the fitting parameter, yields the critical exponent as a function of $t_{\mathrm{max}}$ (Fig.~\ref{f7}).
As can be seen in Fig.~\ref{f7}, the $\nu(t_{\mathrm{max}})$ plots of the $C/T$ results thus obtained approach the value $\nu\,=\,2/3$ of the 3D XY-AFM QCP at low temperatures on both sides of the critical fields, $H_{\mathrm{c1}}$ and $H_{\mathrm{c2}}$, being definitely different from the 2D case $\nu\,=\,1$ and the 3D Ising case $\nu\,=\,0.5$~\cite{RevModPhys.86.563}. The $\nu(t_{\mathrm{max}})$ plots of the $d(T\chi)/dT$ results in Fig.~\ref{f7} support that this behavior of the critical exponent is intrinsic to this system; they also approach the value $\nu\,=\,2/3$ at low temperatures in spite of the disparity of the phase boundaries caused by sample dependence.
In the boson language, the result indicates that the field-induced QCPs of 3-Br-4-F-V belong to the 3D BEC universality class in the temperature range of the present measurements. 
This is the first observation of the 3D BEC exponent in a spin-1/2 ferro-leg ladder, and we reemphasize that there exist few tests of the critical exponent near $H_{\mathrm{c2}}$ in gapped systems, most of which require more than tens of teslas to reach the saturation field and is difficult to examine the critical exponent.

Finally, some remarks are made concerning the relevance of the present study to a BEC in easy-plane ferromagnets in a longitudinal field~\cite{PhysRevB.75.134421,PhysRevB.77.184424}, which has been an attractive topic but less studied experimentally. In the case of strong rung coupling ($|J_{||}/J_\perp|\ll1$) as in 3-Br-4-F-V, the Hamiltonian~(\ref{ladder}) can be mapped onto a spin-1/2 \emph{ferromagnetic} chain with an \emph{easy-plane} anisotropy in an effective magnetic field~\cite{PhysRevB.70.014425}. According to the theoretical study of quasi-one-dimensional ferromagnets with an easy-plane anisotropy (EPFs)~\cite{PhysRevB.75.134421}, it is shown that quasi-one-dimensional EPFs can give a crossover of the critical exponent from the conventional 3D BEC one ($\nu$\,=\,2/3) to $\nu$\,=\,1 as moving away from a field-induced QCP. To observe this crossover, however, interchain couplings must be small enough to satisfy a certain parameter condition to maintain strong one dimensionality. 
In this regard, the coupling parameters of 3-Br-4-F-V might not satisfy the condition to observe this crossover, but the $\nu$\,=\,2/3 behavior at low temperatures agrees with the theory.
Since the other two verdazyl-radical-based FM-leg ladders mentioned above are strong-leg type ($|J_{||}/J_\perp|\,>\,1$)~\cite{PhysRevLett.110.157205,doi:10.7566/JPSJ.83.033707,PhysRevB.91.125104}, they might be the candidates to observe the nontrivial crossover.

\section{Conclusion}
We determined the critical exponents $\nu$ near the field-induced QCPs of the 3D ordering phase boundary on the spin-1/2 ferromagnetic-leg ladder 3-Br-4-F-V, using specific-heat and direct-current magnetization measurements. Near the lower critical field $H_{\mathrm{c1}}$ and the saturation field $H_{\mathrm{c2}}$, the exponents obtained from the temperature-window fitting technique approach the value which belongs to the 3D BEC universality class, $\nu\,=\,2/3$, at low temperatures. Although there is a small sample-dependent difference between the phase boundaries determined from the specific heat and the magnetization, the critical exponents at $H_{\mathrm{c1}}$ and $H_{\mathrm{c2}}$ seem to be unchanged. This fact supports that the 3D BEC exponent is universal in the ordering phase of 3-Br-4-F-V. The verdazyl-radical-based FM-leg ladders have proven to be a model system to study BEC physics.

\section*{Acknowledgments}
This work was supported in part by KAKENHI Grants No. 16J01784, No. 15H03682, No. 17H04850, and No. 15H03695 from JSPS.
The sample preparation of 3-Br-4-F-V was performed at Osaka Prefecture University.
The magnetization and the specific-heat measurements were made at the Institute for Solid State Physics, the University of Tokyo.
Y.~K. is grateful to S. Kittaka for his support in the specific-heat measurements.
%

\end{document}